\documentclass[10pt,conference,compsoc,twocolumn,final,nofonttune]{IEEEtran}

\usepackage{times,amsmath,epsfig}
\usepackage[english]{babel}
\usepackage{amsmath,amsfonts,amsthm} 
\usepackage[hang,small,labelfont=bf,up,textfont=it]{caption}
\usepackage{booktabs} 
\usepackage{lastpage}
\usepackage{graphicx} 

\usepackage{enumitem}
\setlist{noitemsep} 
\usepackage{pstricks}
\usepackage{amssymb}
\usepackage[utf8]{inputenc}
\usepackage{multirow}
\usepackage{amsmath}
\usepackage{mathtools}
\usepackage{tikz}
\usepackage{bclogo}
\usepackage{color, colortbl}
\usepackage{amsthm}
\usepackage{float}
\usepackage{color}

\newcommand{\papertitle}{Toward an Attribute-Based Digital Identity Modelling for Privacy Preservation}

\numberwithin{theorem}{section} 

\usepackage[utf8]{inputenc}
\usepackage{multirow}
\usepackage{graphicx}

\usepackage[T1]{fontenc} 
\usepackage[utf8]{inputenc}

\usepackage{fancyhdr} 
\title{\papertitle}
\author{
{SENE Ibou{\small $~^{\#*1}$}, CISS Abdoul Aziz{\small $~^{\#2}$} and NIANG Oumar{\small $~^{\#3}$} }
\vspace{1.6mm}\\
\fontsize{10}{10}\selectfont\itshape
$^{\#}$\,Ecole Polytechnique de Thiès (E.P.T.), \\Laboratoire de Traitement de l’Information et des Systèmes Intelligents (LTISI),\\
PO Box A10, Thiès, Sénégal\\
\vspace{1.2mm}
\fontsize{10}{10}\selectfont\itshape
$^{*}$\,Université de Thiès (U.T.),\\Ecole Doctorale Développement Durable et Société (ED2DS),\\
PO Box 967, Thiès, Sénégal\\
\fontsize{9}{9}\selectfont\ttfamily\upshape
$^{1}$\,senei@ept.sn,
$^{2}$\,aaciss@ept.sn,
$^{3}$\,oniang@ept.sn
}

\begin{document}
\maketitle

\begin{abstract}
Digital identity is a multidimensional, multidisciplinary, and a complex concept. As a result, it is difficult to apprehend. Plenty of definitions and representations have been proposed so far. However, lots of them are either very generic and difficult to implement in an Attribute-Based Credential context or do not take into account privacy concerns. Seeing how important privacy master is, it becomes a necessity to rethink digital identity in order to take into account privacy concerns. Hence, this paper aims at proposing an Attribute-Based Digital Identity modelling for privacy preservation purposes. The proposed model takes into consideration privacy issues. Thanks to Attribute-Based Credential's secret key, non-transferability and proof of ownership properties, issues about identity theft and security pointed out in many contributions can also be solved.
\end{abstract}


\section{Introduction}
\label{label-introduction}
Physical identities were adequate for face to face (F2F) authentication. However, the Internet is changing everything \cite{20181018291837WIPP} including the way identity is being electronically represented. Indeed, we are living the machine to machine (M2M) period. M2M refers to the communications between computers, embedded processors, smart sensors, actuators, and mobile devices without or with limited human intervention \cite{chen2012machine}. Hence, machines can communicate, exchange data, take decisions without human interventions. Taken decisions  can impact human life in general and their privacy in particular. Therefore, it becomes necessary and urgent to control data disclosed about ourselves to protect our digital identity (DI) in order to master our e-reputation. E-reputation can be considered as an instrumented construction of reputation \cite{domengethal01514313}. It is Therefore the image conveyed or undergone by an entity on the Internet. The term digital identity is said to denote aspects of civil and personal identity that have resulted from the widespread use of identity information to represent people in computer systems \cite{201906FavarinSimone}. If the real identity of an individual is a broad notion that involves psychology, biology, philosophy \cite{vincent:tel-01007682}, etc. how much more for digital identity. Thus, its definition usually depends on usage, situation, purpose and several other factors \cite{phiri2006modelling}. According to the most abstract definition proposed in the literature, digital identity is composed of many aspects including \textit{what one says} (expression), \textit{what one says about others} (opinion), \textit{what one is passionate about} (hobbies), \textit{what one knows} (knowledge), \textit{what represents someone} (avatars), \textit{who one knows} (audience), \textit{what one buys} (consumption), \textit{what is said about somebody} (e-reputation), \textit{what one does} (profession), \textit{what one shares} (publication), etc. From a behavioral point of view, digital identity can be broken down into three components that are \textit{declarative identity} (disclosed data), \textit{acting identity} (activities performed online), and \textit{calculated identity} (resulting from an analysis of acting identity) \cite{2009GeorgesFanny} \cite{georges:hal-01575199}\cite{georges:hal-00332770}. Descriptions provided above point out many beautiful things about digital identity but in practice, it is not as easy as that to take all that aspects into account. P. Daniel was not wrong to assert that "If the identity of organizations is a recognized research theme, its transfer to the digital domain is more problematic" \cite{pelissier:hal-01933420}. Camp JL \cite{camp2004digital} hammers that the lack of conceptual clarity reflected by the overload of the word digital identity is confounding the ubiquitous practice of risk management via identity management. According to Raphael et al. \cite{201903BRPP},  digital identity management is a key issue that will ensure not only service and functionality expectations but also security and privacy.  These few lines show how ambiguous the notion of digital identity is and how difficult it is to pin it down. However, its management is of paramount importance in an increasingly digital world. Even though there are many contributions accordingly, few of them take into account privacy concerns. 

In this paper, we propose an attribute-based digital identity vision usable in an Attribute-Based Credential (ABC) context. ABC is an approach that focuses on attributes of an individual instead of its identifying information and it is known to be the most convenient way to protect user's privacy \cite{isenei2paiotmdpijuin2019}\cite{galparinproceedings}. We refer readers to \cite{isenei2paiotmdpijuin2019} for more details about ABC schemes. The relevance of our contribution lies in the fact that its provides a very simple,  attribute-based model for digital identity modelling and comprehensive enough to be implemented in an ABC context. Also, it helps preventing problem of identity theft and security pointed out in many contributions thanks to ABC credentials' properties. Privacy concerns are taken into account since attributes are never disclosed unless the user decides to do so. The rest of this paper is organized as follows. Related works are presented in section \ref{label-related-works} whereas the proposed model is depicted in section \ref{label-result}. Section \ref{label-conclusion} ends this paper by a conclusion and prospects.

\section{Related works}
\label{label-related-works}

Digital identity is said to be a multidimensional, multidisciplinary, and a complex concept. Its comprehensive apprehension involves many disciplines including philosophy, psychology, biology \cite{vincent:tel-01007682}. Nevertheless, its definition and modelling are very anchored recently. Matthew N. O. Sadiku et al. \cite{2016nosadiku} present a brief introduction to digital identity.  They assert that we live in an age where companies track our digital identities and sell the information to others companies as a form of intellectual property. Their contribution points out how important it is to take care of our digital identity and therefore information we share about ourselves.  J. Pierre \cite{pierre:sic_01084772} directs his contribution towards narrative identity. He focuses on pointing out some challenges namely privacy, security, identity theft, and interoperability. Raphael et al. \cite{201903BRPP} reviewed challenges of identity management systems. They establish an interesting relationship among identities, identifiers and entity. Their representations present some limitations regarding privacy concerns since  they focus on the notion of identity what makes the concerned entity identifiable. Their contribution looks like a procedure of defense against Man in the Middle since they seem to focus on how they can create strong passwords that cannot be easily decoded by man in the middle.  A. Bhargav-Spantzel et al. \cite{squicciarini2005establishing} propose a flexible approach to establish a single sign-on ID in  federation systems. They assert that their contribution is a novel solution for protection against identity theft. However, they do not point out how to formalize DI. Furthermore, their approach depends on Public Key Infrastructure (PKI) for user authentication which requires a trusted third party (Certificate Authority) that knows almost all about everybody. This a threat to privacy preservation. Uciel et al. \cite{fragoso2006federated} present a very interesting vision on relationship between elements of a digital identity. They seem to focus on preserving personal information integrity and confidentiality. However, privacy is deeper than integrity and confidentiality. Their vision of what is a credential is generic since according to them, a credential can be  a password, a certificate, a fingerprint, etc.  Phiri et al. \cite{phiri2006modelling} propose a Digital Identity Management System (DIMS) as a solution for managing digital identity information. The proposed system uses technologies like artificial intelligence and biometrics on the current unsecured networks to maintain the security and privacy of users and service providers in a transparent, reliable and efficient way. Multimode authentication is likely to be the solution for most problems of fraud and identity theft seen on the cyber space today. However, their model may be heavy and difficult to implement in low-resource devices. Armen et al. \cite{khatchatourov:hal-01283997} discuss on pseudonyms and multiple identities. They provide an original analysis grid that can be applied for privacy evaluation in any eID (electronic Identity) architecture. They seem to focus on existing PKI architecture which requires a trusted third party, the so-called Certificate Authority (CA) that knows almost all about everybody. Their vision presents limitations since privacy issues are not taken into account. Grassi et al. \cite{grassi2017digital} present a Digital identity guidelines. However, they seem to focus on describing architecture components and their interrelation instead of depicting DI. Their revision does not explicitly address device identity, often referred to as machine-to-machine authentication. J. Agbinya et al. \cite{200812ajirkc} proposed DEITY (Digital Environment IdentiTY) System for Online Access. Even though their contribution points out many interesting things about DI, lack of privacy consideration may be a limitation. According to their vision, a credential may content all types of metrics especially physical, devices, pseudo, and bio.  

These examples, that can be multiplied, show that, considering digital identity, efforts still have to be made, especially when privacy concerns must be taken into account. 

\section{Result and discussion}
\label{label-result}
\begin{figure}[H] 
\center{
\includegraphics[keepaspectratio=true,scale=0.48]{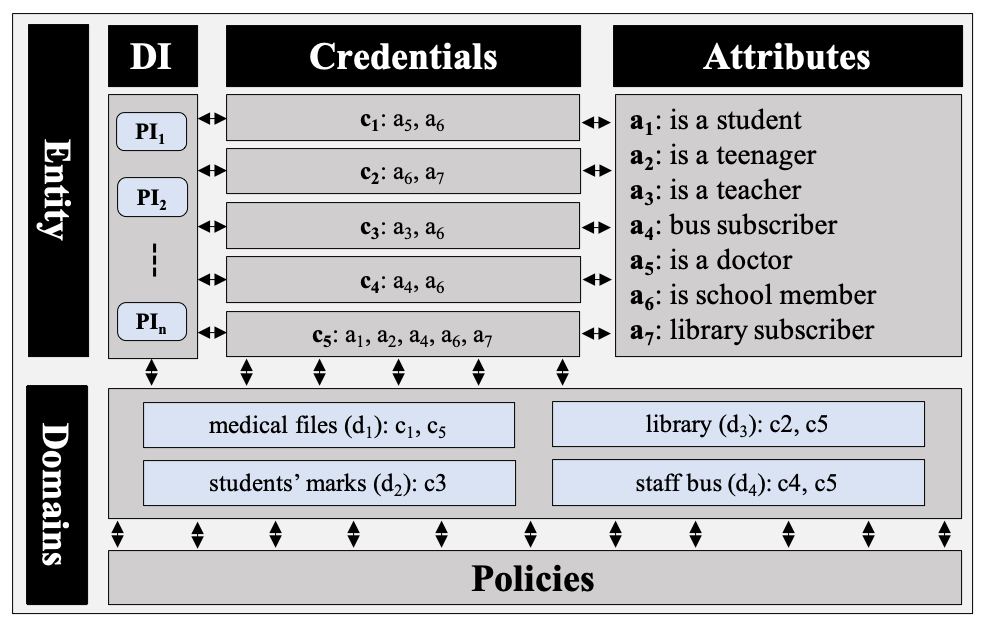} 
\caption{An attribute-based digital identity vision}
\label{fig:di-modelling}
}
\end{figure}

In this section, we present a vision of an attribute-based digital identity. Around this notion, we bring into play some fundamental concepts that are:  attribute, credential, entity, domain, policies and partial identity.  The proposed model is presented in "figure \ref{fig:di-modelling}".

The "figure \ref{fig:di-modelling}" highlights  essential elements involved in the construction of an attribute-based digital identity vision. We detail each of them in the following paragraphs.

\begin{itemize}
\item {
\textbf{Policies}: 
\label{label-policies}
By processing a request of a subject on a resource (object), a system executes a particular function, also known as  operation. Operations can be any of the actions: read, write, edit, delete, create, copy, execute, modify, etc. Policies  define the rules of access to resources as well as the conditions on which those access should be granted. A policy is the description of the rules that determine the permitted operations that a subject can perform on a resource (object) under specific conditions. In short, policy refers to a set of directives, also known as rules, that specify who has permission to do what on which resource and under what conditions. This definition highlights four concepts that are:

\begin{itemize}
\item {
\textbf{Subject}: It is the entity that wishes to perform an action on the requested resource.
}
\item {
\textbf{Object: } It is the resource on which the requested action should be performed.
}
\item {
\textbf{Action:} it corresponds to the function or operation that must be performed on the requested resource.
}
\item {
\textbf{Context:} It determines the characteristics of the query's execution environment and the underlying conditions.
}
\end{itemize}

Policies may be transversal to several domains. Nevertheless, domains do not necessarily have the same policies.
}

\item {
\textbf{Domain}: 
\label{label-domain} We introduce  the concept of domain and consider that it is a set of resources governed by the same policies. Access to a domain requires authentication. Thus, accessing to the domain of the library requires, for example, a qualification to be a subscriber of that library whereas access to students' marks requires a predisposition to be a teacher. Consequently, a domain is a set of resources  on which it is possible to perform operations according to  well-defined  policies. In the domain of the library for instance, an entity must be able to request the action  read on a book resource  whereas  in the medical files domain, an entity may request the action write on a patient file resource.
}

\item {
\textbf{Entity}: 
\label{label-entity} According to the computer and internet dictionary, an entity is any concrete or abstract object, having or not an existence of its own, that is to say that can be described or manipulated without requiring knowledge of other entities. In general, an entity is an object with characteristics also known as attributes.  In our case, we focus on entities of type subject described in section  \ref{label-policies}.  Therefore, an entity refers to a subject that should request operations on  resources.
}

\item {
\textbf{Attribute}:
\label{label-definition-attribute} An attribute is a characteristic of an entity. It can be static, dynamic, innate, temporary, etc. An attribute is composed of two parts that are its name and its value \cite{afshar2018attribute}.  It can describe almost anything: who, what, when, how, where, for what, etc. An attribute certified as compliant by an official or a trusted third party is known as "claim" \cite{harry2013iam}. In a Senegalese context for instance, it is required to have the characteristic "over 18 years" to be able to play some gambles. Attributes are indirectly  transversal to partial identities and domains across credentials. 
}

\item {
\textbf{Credential}:
\label{label-definition-credential} Historically, especially in the Middle Ages, a credential was a letter handed to a traveler by a king, an important noble or an ecclesiastical authority and which established his names and qualities and allowed him to be received by relatives, friends or debtors of the signatory. In the religious point of view, the pilgrim wishing to go to a distant sanctuary asked for a credential, a document attesting his state of pilgrimage commendable to those who could offer him hospitality. It also served him as a pass to the civil, military and ecclesiastical authorities he met along the way. Even if its transposition in the digital world changes its format, it goes without saying that its objective remains the same. Credentials attest that an entity has a certain knowledge, skill, characteristic, etc. \cite{isenei2paiotmdpijuin2019}\cite{de2017assessment}. They involve attributes of an entity without including identity information, which allows linking the credential to its owner \cite{de2017assessment}\cite{lueks2017fast}\cite{de2018attribute}\cite{isenei2paiotmdpijuin2019}. So, a credential can be used as many times as necessary without saying more about its holder. This is one of the flagship properties of an ABC: multi show unlinkability. As illustrated in "figure \ref{fig:credential}", a credential has four main parts that are: a secret key of its owner, a set of claims, a signature of a trusted third party also known as issuer, and metadata of the credential.

\begin{figure}[H] 
\center
\includegraphics[keepaspectratio=true,scale=0.65]{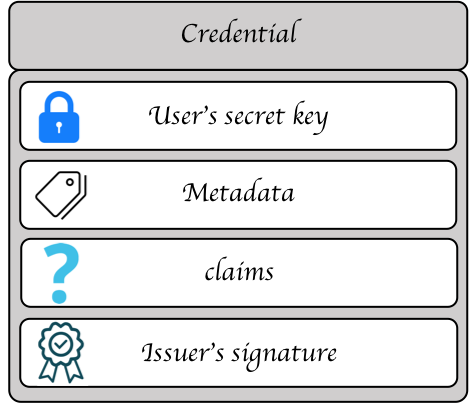} 
\caption{Credential's structure.}
\label{fig:credential}
\end{figure}
}

\item {
\textbf{Partial identity}: 
\label{label-partial-identity} In the digital area, an entity is not seen in the same way depending on whether it is in a domain of e-commerce, leisure, governmental, professional, health, etc. These changes of identity depending on the situation are represented by partial identities \cite{200812ajirkc}. An entity interacts differently with each domain and so each will have a different picture of “who it  is” and “what it does.” The combination of those  partial identities makes up “who it is and what it does” \cite{200812ajirkc}. A partial identity can be seen as a subset of the characteristics of an entity that make up its identity in a particular domain.
}

\item {
\textbf{Digital identity}:
 \label{label-digital-identity} The definition of digital identity (DI) is an area that has been well anchored in recent years. According to Sadiku et al. \cite{2016nosadiku}, a DI is the digital representation of the information on a person, organization or object. It describes the virtual identity of a user in a computer network. Kim \cite{cameron2005laws} defines DI as a set of claims made by one digital subject about itself or another digital subject.  A claim is an assertion of the truth of something, typically one which is disputed or in doubt. In the words of Fragoso et al. \cite{fragoso2006federated}, DI can be considered as the electronic representation of an entity within a domain of application. By Paul et al. \cite{grassi2017digital}, without context, it is difficult to land on a single definition that satisfies all requirements. They consider DI as the online persona of a subject.  Sittampalam \cite{subenthiran2005digital} defines DI as a virtual representation of a real identity that can be used in electronic interactions with other machines or people.  These  definitions, which can be multiplied, all have a common denominator. Indeed, a digital identity links two types of entities: a real entity and a virtual one.  So as far as we are concerned, we consider  digital identity as a set of claims subject to doubt  about an entity. This set of claims is generally divided into subset  known as partial identities.  The "figure \ref{fig:pid-id}" is an illustration of partial  and digital identities inspired form \cite{hansen2008identity}.

\begin{figure}[H] 
\center
\includegraphics[width=0.5\textwidth]{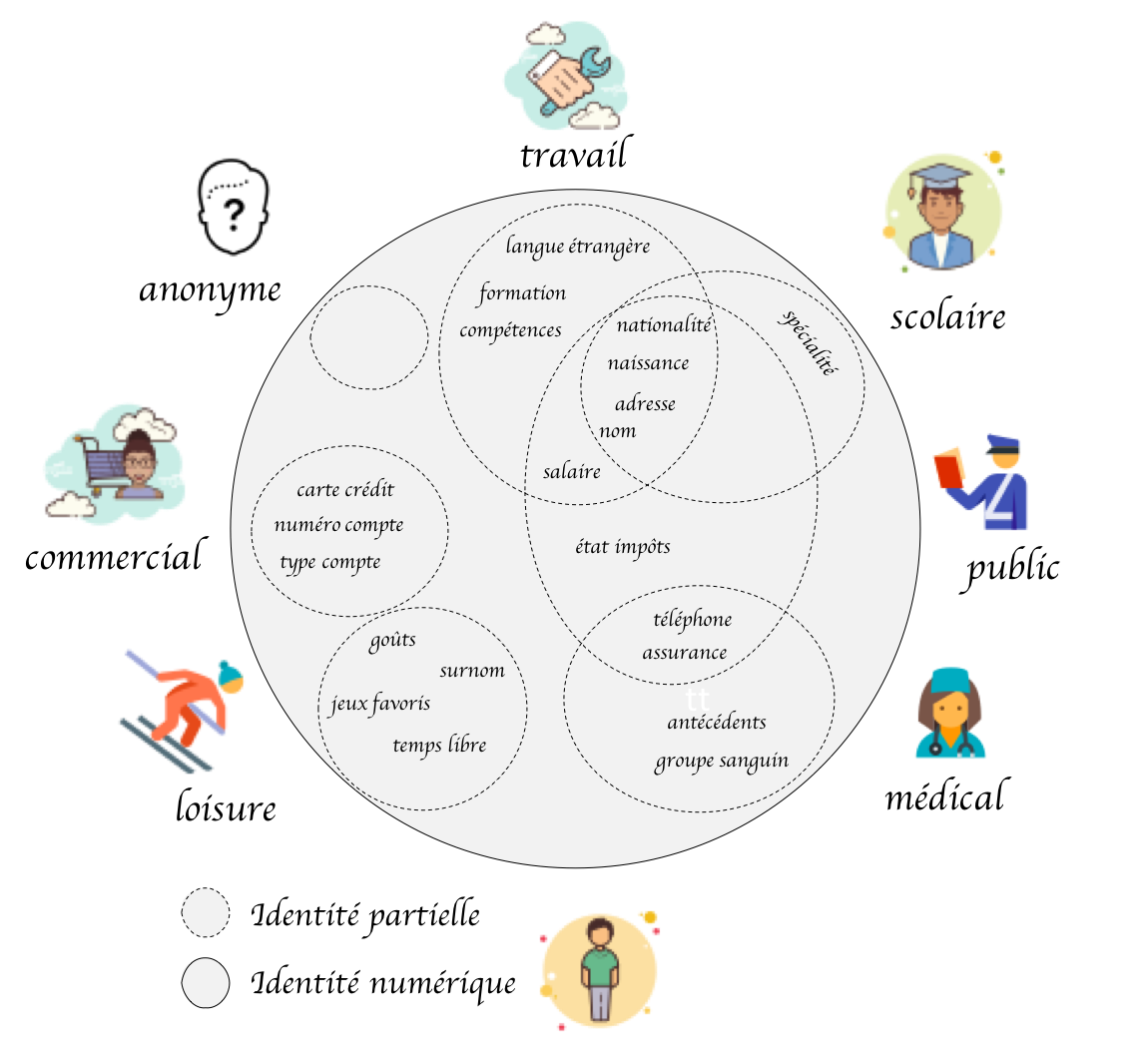} 
\caption{Partial identities and digital identity }
\label{fig:pid-id}
\end{figure}
}
\end{itemize}

We have just described the  elements involved in the modelling of  an attribute-based digital identity. Attributes are the base of our model. They are transversal to domains and partial identities across credentials. A credential can be transversal to several domains and domains can require presentation of one or more credentials. If we assume that access to \textit{medical files} (respectively students' marks, library, staff bus) requires attributes $a_5$ and $a_6$ (respectively $a_3$, $a_6$ and $a_7$, $a_4$ and $a_6$ )  then credentials  $c_1$ and $c_5$ (respectively $c_3$, $c_2$ and $c_5$, $c_4$ and $c_5$) grant access to  domain $d_1$( respectively $d_2$,$d_3$, and $d_4$). So, an example of textual policy should be: \textit{only students who are  school member and library subscribers must be able to read library's audio file and this between 8am and 6pm except the weekend.} By deconstructing this policy, we note that:

\begin{itemize}
\item {
Subjects with attributes $a_1,$ $a_6,$ $a_7$  are targeted;
}
\item {
The concerned resources  are those of  type audio;
}
\item {
The action that should be executed is "read";
}
\item {
The context highlights a temporal aspect by limiting the hours of access as well as the days of the week;
}
\item {
The "library" domain is the targeted one;
}
\end{itemize}

\section{Conclusion}
\label{label-conclusion}

The real identity was adapted in face-to-face authentication. Nevertheless, its transfer to the digital world has caused lots of challenges including issues about privacy preservation. In his paper, we propose an attribute-based digital identity modelling to take into account privacy issues. The proposed  model takes into consideration three fundamental aspects, namely security, privacy and identity theft. It can be  implemented in an ABC context. However, for a successful deployment,  attributes must be standardized to allow organizations and practitioners to speak the same language. This would make it possible to define attribute-based policies (ABP) understandable by everybody.

\bibliographystyle{IEEEtran}
\bibliography{IEEEabrv,IEEEexample}

\begin{thebibliography}{10}
\providecommand{\url}[1]{#1}
\csname url@samestyle\endcsname
\providecommand{\newblock}{\relax}
\providecommand{\bibinfo}[2]{#2}
\providecommand{\BIBentrySTDinterwordspacing}{\spaceskip=0pt\relax}
\providecommand{\BIBentryALTinterwordstretchfactor}{4}
\providecommand{\BIBentryALTinterwordspacing}{\spaceskip=\fontdimen2\font plus
\BIBentryALTinterwordstretchfactor\fontdimen3\font minus
  \fontdimen4\font\relax}
\providecommand{\BIBforeignlanguage}[2]{{%
\expandafter\ifx\csname l@#1\endcsname\relax
\typeout{** WARNING: IEEEtran.bst: No hyphenation pattern has been}%
\typeout{** loaded for the language `#1'. Using the pattern for}%
\typeout{** the default language instead.}%
\else
\language=\csname l@#1\endcsname
\fi
#2}}
\providecommand{\BIBdecl}{\relax}
\BIBdecl

\bibitem{20181018291837WIPP}
W.~Inambao, J.~Phiri, and D.~Kunda, ``Digital identity modelling for digital
  financial services in zambia,'' \emph{ICTACT Journal on Communication
  Technology}, vol.~9, no.~3, pp. 1829--1837, 10 2018.

\bibitem{chen2012machine}
M.~Chen, J.~Wan, and F.~Li, ``Machine-to-machine communications: Architectures,
  standards and applications.'' \emph{Ksii transactions on internet \&
  information systems}, vol.~6, no.~2, 2012.

\bibitem{domengethal01514313}
\BIBentryALTinterwordspacing
J.-C. DOMENGET, ``{Construire son e-r{\'e}putation sur Twitter},'' in
  \emph{{E-r{\'e}putation : regards crois{\'e}s sur une notion {\'e}mergente}},
  C.~Alcantara, Ed.\hskip 1em plus 0.5em minus 0.4em\relax {Lextenso
  {\'e}ditions}, Mar 2015, pp. 135--143. [Online]. Available:
  \url{https://hal.archives-ouvertes.fr/hal-01514313}
\BIBentrySTDinterwordspacing

\bibitem{201906FavarinSimone}
S.~Favarin, ``How design thinking helps digital identity design,'' 06 2019.

\bibitem{vincent:tel-01007682}
\BIBentryALTinterwordspacing
J.~Vincent, ``{Digital Identity for a Telecom Operator},'' Theses,
  {Universit{\'e} de Caen}, Jun. 2013. [Online]. Available:
  \url{https://tel.archives-ouvertes.fr/tel-01007682}
\BIBentrySTDinterwordspacing

\bibitem{phiri2006modelling}
J.~Phiri and J.~I. Agbinya, ``Modelling and information fusion in digital
  identity management systems,'' in \emph{Networking, International Conference
  on Systems and International Conference on Mobile Communications and Learning
  Technologies, 2006. ICN/ICONS/MCL 2006. International Conference on}.\hskip
  1em plus 0.5em minus 0.4em\relax IEEE, 2006, pp. 181--181.

\bibitem{2009GeorgesFanny}
F.~Georges, ``Représentation de soi et identité numérique,''
  \emph{Réseaux}, vol. 154, pp. 165--193, 08 2009.

\bibitem{georges:hal-01575199}
\BIBentryALTinterwordspacing
------, ``L'identit{\'e} num{\'e}rique dans le web 2.0,'' Mar 2008. [Online].
  Available: \url{https://hal.archives-ouvertes.fr/hal-01575199}
\BIBentrySTDinterwordspacing

\bibitem{georges:hal-00332770}
\BIBentryALTinterwordspacing
------, ``Les composantes de l'identit{\'e} dans le web 2.0, une {\'e}tude
  s{\'e}miotique et statistique. hypostase de l'imm{\'e}diatet{\'e},'' in
  \emph{{Communication au 76{\`e}me congr{\`e}s de l'ACFAS: Web participatif:
  mutation de la communication ?, 6 et 7 mai 2008, Centre des congr{\`e}s,
  Qu{\'e}bec.}}, Qu{\'e}bec, Canada, May 2008, p.~12. [Online]. Available:
  \url{https://hal.archives-ouvertes.fr/hal-00332770}
\BIBentrySTDinterwordspacing

\bibitem{pelissier:hal-01933420}
\BIBentryALTinterwordspacing
D.~P{\'e}lissier, ``{Identit{\'e} num{\'e}rique des organisations : approche
  conceptuelle et analyse exploratoire de r{\'e}ception par la
  lexicom{\'e}trie},'' in \emph{{Colloque Jeunes Chercheurs Praxiling 2015,
  Trace(s)}}, Montpellier, France, Oct. 2015. [Online]. Available:
  \url{https://hal.archives-ouvertes.fr/hal-01933420}
\BIBentrySTDinterwordspacing

\bibitem{camp2004digital}
J.~Camp, ``Digital identity,'' \emph{IEEE Technology and society Magazine},
  vol.~23, no.~3, pp. 34--41, 2004.

\bibitem{201903BRPP}
R.~Banda and P.~Phiri, ``Challenges of identity management systems and
  mechanisms: A review of mobile identity,'' 03 2019.

\bibitem{isenei2paiotmdpijuin2019}
I.~Sene, A.~Ciss, and O.~Niang, ``I2pa: An efficient abc for iot,''
  \emph{Cryptography}, vol.~3, no.~2, p.~16, 06 2019.

\bibitem{galparinproceedings}
G.~Alpar and J.-H. Hoepman, ``A secure channel for attribute-based credentials:
  [short paper],'' 11 2013, pp. 13--18.

\bibitem{2016nosadiku}
M.~N~O~Sadiku, A.~Shadare, and S.~M~Musa, ``Digital identity,''
  \emph{International Journal of Innovative Science, Engineering and
  Technology}, vol.~3, p. 2016, 12 2016.

\bibitem{pierre:sic_01084772}
\BIBentryALTinterwordspacing
J.~Pierre, ``{De l'identit{\'e} num{\'e}rique {\`a} l'individu
  transm{\'e}diatique},'' \emph{{MEDIADOC}}, no.~13, Dec. 2014. [Online].
  Available: \url{https://archivesic.ccsd.cnrs.fr/sic_01084772}
\BIBentrySTDinterwordspacing

\bibitem{squicciarini2005establishing}
A.~Bhargav-Spantzel, A.~C. Squicciarini, and E.~Bertino, ``Establishing and
  protecting digital identity in federation systems,'' \emph{Journal of
  Computer Security}, vol.~14, no.~3, pp. 269--300, 2006.

\bibitem{fragoso2006federated}
U.~Fragoso-Rodriguez, M.~Laurent-Maknavicius, and J.~Incera-Dieguez,
  ``Federated identity architectures,'' in \emph{Proc. 1st Mexican Conference
  on Informatics Security 2006 (MCIS’2006)}, 2006.

\bibitem{khatchatourov:hal-01283997}
\BIBentryALTinterwordspacing
A.~Khatchatourov, M.~Laurent, and C.~Levallois-Barth, ``{Privacy in digital
  identity systems: models, assessment and user adoption},'' in \emph{{14th
  International Conference on Electronic Government (EGOV)}}, ser. Electric
  government, E.~Tambouris, M.~Janssen, H.~J. Scholl, M.~A. Wimmer,
  K.~Tarabanis, M.~Gasc{\'o}, B.~Klievink, I.~Lindgren, and P.~Parycek, Eds.,
  vol. LNCS-9248.\hskip 1em plus 0.5em minus 0.4em\relax Thessaloniki, Greece:
  {Springer international publishing}, Aug. 2015, pp. 273--290, part 4:
  Application Areas and Evaluation. [Online]. Available:
  \url{https://hal.archives-ouvertes.fr/hal-01283997}
\BIBentrySTDinterwordspacing

\bibitem{grassi2017digital}
P.~A. Grassi, M.~E. Garcia, and J.~L. Fenton, ``Digital identity guidelines,''
  \emph{NIST special publication}, vol. 800, pp. 63--3, 2017.

\bibitem{200812ajirkc}
J.~Agbinya, R.~Islam, and C.~Kwok, ``Development of digital environment
  identity (deity) system for online access,'' 12 2008, pp. 1 -- 8.

\bibitem{afshar2018attribute}
M.~Afshar, S.~Samet, and T.~Hu, ``An attribute based access control framework
  for healthcare system,'' in \emph{Journal of Physics: Conference Series},
  vol. 933, no.~1.\hskip 1em plus 0.5em minus 0.4em\relax IOP Publishing, 2018,
  p. 012020.

\bibitem{harry2013iam}
G.~Harry, ``Iam-gestion des identit{\'e}s et des acc{\`e}s: concepts et
  {\'e}tats de l'art,'' 2013.

\bibitem{de2017assessment}
J.~De~Fuentes, L.~Gonz{\'a}lez-Manzano, J.~Serna-Olvera, and F.~Veseli,
  ``Assessment of attribute-based credentials for privacy-preserving road
  traffic services in smart cities,'' \emph{Personal and Ubiquitous Computing},
  vol.~21, no.~5, pp. 869--891, 2017.

\bibitem{lueks2017fast}
W.~Lueks, G.~Alp{\'a}r, J.-H. Hoepman, and P.~Vullers, ``Fast revocation of
  attribute-based credentials for both users and verifiers,'' \emph{Computers
  \& Security}, vol.~67, pp. 308--323, 2017.

\bibitem{de2018attribute}
J.~M. de~Fuentes, L.~Gonzalez-Manzano, A.~Solanas, and F.~Veseli,
  ``Attribute-based credentials for privacy-aware smart health services in
  iot-based smart cities,'' \emph{Computer}, vol.~51, no.~7, pp. 44--53, 2018.

\bibitem{cameron2005laws}
K.~Cameron, ``The laws of identity,'' \emph{Microsoft Corp}, vol.~12, pp.
  8--11, 2005.

\bibitem{subenthiran2005digital}
S.~Subenthiran, ``Digital identity modelling and management,'' Ph.D.
  dissertation, 2005.

\bibitem{hansen2008identity}
M.~Hansen, A.~Pfitzmann, and S.~Steinbrecher, ``Identity management throughout
  one's whole life,'' \emph{Information security technical report}, vol.~13,
  no.~2, pp. 83--94, 2008.

\end{thebibliography}
\end{document}